\documentclass[preprint,aps,showpacs,nofootinbib,tightenlines]{revtex4-1}
\usepackage{amsmath}
\usepackage{amssymb}
\usepackage{epsfig}
\usepackage{graphicx}
\begin{document}
%%%%%%%%%%%%%%%%%%%%%%%%%%%%%%%%%%%%%%%%%%%%%

\newcommand{\beq}{\begin{eqnarray}}
\newcommand{\eeq}{\end{eqnarray}}
\newcommand{\non}{\nonumber\\ }

\newcommand{\ov}{\overline}

\newcommand{\psl}{ P \hspace{-2.5truemm}/ }
\newcommand{\nsl}{ n \hspace{-2.2truemm}/ }
\newcommand{\vsl}{ v \hspace{-2.2truemm}/ }
\newcommand{\epsl}{\epsilon \hspace{-1.6truemm}/\,  }

\def \epjc{ Eur. Phys. J. C }
\def \jpg{  J. Phys. G }
\def \npb{  Nucl. Phys. B }
\def \plb{  Phys. Lett. B }
\def \pr{  Phys. Rep. }
\def \rmp{ Rev. Mod. Phys. }
\def \prd{  Phys. Rev. D }
\def \prl{  Phys. Rev. Lett.  }
\def \zpc{  Z. Phys. C  }
\def \jhep{ J. High Energy Phys.  }
\def \ijmpa { Int. J. Mod. Phys. A }

%%%%%%%%%%%%%%%%%%%%%%%%%%%%%%%%%%%%%%%%%%%%%%%%%%%%
%%
\title{Revisiting the pure annihilation decays $B_s\to \pi^+ \pi^-$ and
$B^0 \to K^+ K^-$: the data and the pQCD predictions}
\author{Zhen-Jun Xiao\footnote{xiaozhenjun@njnu.edu.cn}, Wen-Fei Wang and Ying-Ying Fan }
\affiliation{
Department of Physics and Institute of Theoretical Physics,\\
Nanjing Normal University, Nanjing, Jiangsu 210046, People's Republic of China}
\date{\today}
\begin{abstract}
In this work, we recalculate the charmless pure annihilation decays
$B_s\to \pi^+ \pi^-$  and $B^0 \to K^+ K^-$ by using the perturbative QCD
(pQCD) factorization approach, and compare the pQCD predictions with
currently available experimental measurements.
By numerical calculations and phenomenological analysis we found the following results:
(a) one can provide a consistent pQCD interpretation for both the
measured $Br(B_s^0 \to \pi^+ \pi^-)$ and $Br(B_d^0 \to K^+ K^-)$ simultaneously;
(b) the pQCD predictions for $Br(B_s^0 \to \pi^+\pi^- )$ obtained by different authors are well consistent
with each other;
(c) our new pQCD prediction for $Br(B_d^0 \to K^+K^- )$ agree well with the measured values
from CDF and LHCb Collaboration;
and (d) the CP-violating asymmetry ${\cal A}_{CP}(B_d^0 \to K^+K^-) \approx 19\%$, which is large and
may be detected at the LHCb and future Super-B factory experiments.
\end{abstract}

\pacs{13.25.Hw, 12.38.Bx, 14.40.Nd}

\maketitle

\section{Introduction}\label{sec:1}

Among the two-body hadronic B meson decays, the pure annihilation decay modes,
such as $B_s^0 \to \pi^+\pi^-$ and $B^0 \to K^+ K^-$ decays,  are specific in several respects.
They can occur only through the annihilation diagrams in the standard
model (SM) because none of the quarks(anti-quarks)
in the final states are the same as those of the initial B meson. And consequently, they
are rare decay modes with a branching ratio at the level of $10^{-7}$ or less as generally expected.
Such decays play very important role in understanding the annihilation mechanism and determining the strength of the
annihilation contribution in B meson charmless hadronic decays, and therefore have been studied intensively
by many authors\cite{chen2001,liy2004,ali2007,xiao2008,bn675,duprd68,yang2005,cheng1,cheng2,zhu2011} in spite of the great difficulties
in both the theoretical calculation and the experimental measurements.

In the experiment side, both $Br(B_s^0 \to \pi^+ \pi^-)$ and $Br(B^0 \to K^+ K^-)$  are measured very recently
due to their rareness. At the spring and summer conference of 2011, CDF \cite{cdf-2011a} and LHCb \cite{lhcb-2011a} collaboration reported their first measurement of the decay rates
\beq
Br(B_s^0 \to \pi^+\pi^- )&=&
\left\{ \begin{array}{ll} ( 5.7 \pm 1.5 (stat.) \pm 1.0 (syst.) ) \times 10^{-7} &{\rm CDF}\quad [11]\\
(9.8^{+2.3}_{-1.9}(stat.) \pm 1.1 (syst.) ) \times 10^{-7}, & {\rm LHCb}\ \  [12],\\ \end{array} \right.
\label{eq:exppi}\\
Br(B^0 \to K^+K^- )&=& \left\{ \begin{array}{ll}
(2.3\pm 1.0(stat.) \pm 1.0 (syst.))\times 10^{-7}, &  {\rm CDF}\ \ [11], \\
(1.3^{+0.6}_{-0.5}(stat.)\pm 0.7(syst.))\times 10^{-7}, &  {\rm LHCb}\ \ [12]. \\
\end{array} \right.
\label{eq:expk}
\eeq
The statistical significance of LHCb measurement reaches $5.3\sigma$ for $B_s\to \pi^+\pi^-$ decay, which means  a observation for the first time.

In the theory side, we know that it is very hard to make a reliable calculation for pure annihilation decays of B mesons.
In the QCD factorization (QCDF) approach\cite{bbns1999}, for example, one can not perform a real calculation
for the annihilation diagrams due to the end-point singularity, but have to make an rough
estimation by parameterizing the annihilation contribution through the treatment
$\int^{1}_{0}dx/x \to X_A=(1+\rho_A e^{i\phi})\ln\frac{m_B}{\Lambda_h}$ \cite{bn675,duprd68}, or by using
an effective gluon propagator $1/k^2 \to 1/\left ( k^2+M_g^2(k^2)\right )$ to avoid enhancements in the soft endpoint region
\cite{yang2005}. Of course, such parameterization will produce large theoretical uncertainties.
For $B_s^0 \to \pi^+\pi^-$ and $B^0 \to K^+ K^-$ decays, the theoretical predictions based on the QCDF approach
as given for example in Refs.\cite{bn675,duprd68,yang2005,cheng1,cheng2}  are the following:
\beq
Br(B_s^0 \to \pi^+\pi^- )&=& \left\{ \begin{array}{ll}
 0.24 \times 10^{-7}, & [5,6] \\ (1.24\pm 0.28)\times 10^{-7}, & [7] \\
(2.6\pm 1.0)\times 10^{-7}, & [9]\\
\end{array} \right. \label{eq:qcdfpi}
\\
Br(B^0 \to K^+K^- )&=& \left\{ \begin{array}{ll} 0.13 \times 10^{-7}, & [5] \\
(1.0^{+0.3}_{-0.2}\pm 0.3)\times 10^{-7}, & [8]\\
\end{array} \right. . \label{eq:qcdfk}
\eeq
Obviously, the QCDF predictions in Refs.~\cite{bn675,duprd68,yang2005,cheng1} are much smaller than the measured
results for $Br(B_s^0 \to \pi^+\pi^- )$, while smaller or close to the measured ones for $Br(B^0 \to K^+K^- )$
in Ref.~\cite{bn675} and Ref.~\cite{cheng1}, respectively.

After CDF's report of the evidence of $B_s \to \pi^+\pi^-$ decay, the author of
Ref.~\cite{zhu2011} reinvestigated
the role of annihilation topology in the QCDF approach and found that (1) the CDF
measurement of
$Br(B_s\to \pi^+\pi^-)$ implies a large annihilation scenario with $\rho_A$
around $2$ instead of $\rho_A\approx 1$ preferred by
all previous studies in QCDF approach\cite{bn675,duprd68,bbns1999};
(2) if one assumes universal annihilation
parameters $\rho_A$ and $\phi_A$ for  all $B_{d,s} \to PP$ decay modes,
one can not provide predictions being
consistent with all well measured decays\footnote{The corresponding QCDF
predictions for $Br(B_s\to K^+K^-)$ and
$Br(B_d\to K^0\bar{K}^0)$ are twice larger than the experimental
measurements \cite{zhu2011}.}; (3) one possible way to solve
this problem is to use different $(\rho_A,\phi_A)$ for different
decays, which however means that the predictive power
of QCDF approach becomes rather limited. In short the studies in
Ref.~\cite{zhu2011} tell us that it is very
hard to give a consistent
QCDF interpretation for $Br(B_s\to \pi^+\pi^-)$ and other well measured
$B_{d,s} \to PP$ decay modes simultaneously.

In the perturbative QCD (pQCD) factorization approach \cite{keum01-a,keum01-b,lu01:pipi,li2003},
however, the situation
becomes rather different. Here, the pure annihilation decays of $B/B_s$
meson can be calculated perturbatively
by employing the
Sudakov factors to smear and then to strongly suppress the end-point singularity.
In the pQCD factorization approach,
for example, the endpoint divergence of the factorizable emission diagram
Fig.1(a) and 1(b) in Ref.~\cite{wang2006a}
are regulated by introducing the transverse momentum $k_{i\perp}^2$, i.e.
\beq
\frac{1}{k^2}\cdot \frac{1}{p_b^2-M_B^2}&=&\frac{1}{M_B^4 x_1 x_3 (1-x_3)} \non
&\longrightarrow &
\frac{1}{(1-x_3)M_B^2 + {\bf k_{3\perp}^2}} \cdot \frac{1}{x_1x_3M_B^2
+ ({\bf k_{1\perp}^2}-{\bf k_{3\perp}^2})^2},
\eeq
where $k^2=(k_1-k_3)^2$ and $p_b^2=(P_1-k_3)^2$ is the momentum of the gluon propagator
and $b$-quark propagator respectively. It is easy too see that the end-point divergence
for $x_1=0$ and $x_3=(0,1)$ are removed effectively by introducing small but non-zero
${\bf k_{i\perp}^2}$.

For $B_s^0 \to \pi^+ \pi^-$ decay, it was calculated by employing the pQCD
factorization approach in 2004 \cite{liy2004} and 2007 \cite{ali2007}, respectively.
In Ref.~\cite{liy2004}, we obtained the first pQCD prediction for the decay rate:
\beq
Br(B_s^0 \to \pi^+\pi^- )=(4.2\pm 0.6)\times 10^{-7}.
\eeq
In 2007, Ali {\it et al.,} \cite{ali2007} made a systematic calculation
for all $B_s \to PP, PV, VV$ decays
in the pQCD factorization approach and found that
\beq
Br(B_s^0 \to \pi^+\pi^- )=(5.7^{+1.8}_{-1.6})\times 10^{-7}.
\eeq
These two pQCD predictions at leading order (LO) are well consistent within
$1\sigma$ error and confirmed by CDF and LHCb measurements as shown in
Eqs.~(\ref{eq:exppi},\ref{eq:expk}).
The small difference for the predicted decay rates between Ref.~\cite{liy2004}
and \cite{ali2007} comes from the fact that
a little different input parameters and distribution amplitudes(DA's)
of $\pi$ and $B_s$ meson were used in two studies.

In Ref.~\cite{xiao2008}, by employing the pQCD factorization approach,
we studied the $B_s \to PP$ decays with the inclusion
of partial next-to-leading order (NLO) contributions, coming from the QCD vertex
corrections,
the quark-loops, the chromo-magnetic penguins and the usage of the NLO Wilson
coefficients instead of the LO ones.
For the pure annihilation decay $B_s \to \pi^+ \pi^-$, it does not receive the
 NLO  contributions from the QCD vertex corrections, the quark-loops and the
 chromo-magnetic penguins. The leading order pQCD prediction is $Br(B_s^0
 \to \pi^+\pi^-)= (7\pm 2.5)
\times 10^{-7}$, while it becomes $(5.7^{+2.4}_{-2.2})\times 10^{-7}$
when the NLO Wilson coefficients $C_i(M_W)$, the NLO renormalization
group evolution matrix $U(t,m,\alpha)$ \cite{buras96} and the $\alpha_s(t)$
at two-loop level were employed in the numerical calculation \cite{xiao2008}.

For $B^0 \to K^+ K^-$ decay, the known pQCD prediction for its branching ratio
was given in 2001 \cite{chen2001}
\beq
Br(B_d^0 \to K^+K^- )= 3.27 \times 10^{-8}; \qquad Br(\bar{B}_d^0 \to K^+K^- )=
 5.90 \times 10^{-8},
\eeq
which is much smaller than the measured value as given in Eq.~(\ref{eq:expk})
by roughly a factor of three, in other
words, a large discrepancy between the data and the theoretical prediction
based on the pQCD factorization approach
for $B^0 \to K^+ K^-$ decay.

It is necessary and interesting to check if one can provide a consistent pQCD
interpretation for both the
measured $Br(B_s^0 \to \pi^+ \pi^-)$ and $Br(B_d^0 \to K^+ K^-)$ simultaneously?
In this paper, by employing the pQCD factorization approach, we recalculate the
pure annihilation decays $B_s^0 \to \pi^+\pi^-$ and $B^0\to K^+ K^-$ with the
usage of the same set of input parameters
and wave functions for the mesons involved, in order to check if the new data
from CDFF and LHCb can be understood in the pQCD approach.
Our studies will be helpful to determine the strength of penguin-annihilation
amplitudes \cite{buras697}.

The paper is organized as follows. In Sec.~\ref{sec:2}, we give a brief
review about the
theoretical framework of the pQCD factorization approach and the wave functions
for $B^0/B_s^0$ and $\pi, K$ mesons involved.
We perform the perturbative calculations for considered decay channels
in Sec.~\ref{sec:3},
while the numerical results and phenomenological analysis are given in
Sec.~\ref{sec:4}. A short summary also be given in Sec.~\ref{sec:4}.

\section{Theoretical Framework} \label{sec:2}

In the pQCD approach, the decay amplitude ${\cal A}(B_q \to M_2 M_3)$ with $q=(d,s)$
can be written conceptually as the convolution,
\beq
{\cal A}(B_q \to M_2 M_3)\sim \int\!\! d^4k_1 d^4k_2 d^4k_3\ \mathrm{Tr}
\left [ C(t) \Phi_B(k_1) \Phi_{M_2}(k_2) \Phi_{M_3}(k_3)H(k_1,k_2,k_3, t) \right ],
\label{eq:con1}
\eeq
where $k_i$'s are momenta of light quarks included in each meson, and $``\mathrm{Tr}"$
denotes the trace over Dirac and color indices.
In the above convolution, $C(t)$ is the Wilson coefficient evaluated at scale $t$,
the function $H(k_1,k_2,k_3,t)$ describes the four
quark operator and the spectator quark connected by  a hard gluon.
The wave function $\Phi_B(k_1)$ and $\Phi_{M_i}$ describe the hadronization of the
quark and anti-quark in the $B_q$ meson and the final state light meson $M_i$.

We treat the $B_q$ meson as a heavy-light system, and consider the $B_q$ meson at rest
for simplicity. Using the light-cone coordinates the $B_q$ meson momentum $P_B$ and the two
final state meson's momenta $P_2$ and $P_3$ (for $M_2$ and $M_3$ respectively)
can be written as
\beq
P_B = \frac{M_B}{\sqrt{2}} (1,1,{\bf 0}_{\rm T}), \quad
P_2 = \frac{M_B}{\sqrt{2}}(1-r_3^2,r^2_2,{\bf 0}_{\rm T}), \quad
P_3 = \frac{M_B}{\sqrt{2}} (r_3^2,1-r^2_2,{\bf 0}_{\rm T}),
\eeq
where $r_i=m_i/M_B$. For the final state light mesons made up with $(u,d,s)$ and the corresponding anti-quarks,
the ratio $r_2$ and $r_3$ are small and will be neglected safely. Putting the quark momenta in $B_q$, $M_2$
and $M_3$ meson as $k_1$, $k_2$, and $k_3$, respectively, we can choose
\beq
k_1 = (x_1 P_1^+,0,{\bf k}_{\rm 1T}), \quad
k_2 = (x_2 P_2^+,0,{\bf k}_{\rm 2T}), \quad
k_3 = (0, x_3 P_3^-,{\bf k}_{\rm 3T}).
\eeq
Then, the integration over $k_1^-$, $k_2^-$, and $k_3^+$ in eq.(\ref{eq:con1}) will lead to
\beq
{\cal A}(B_q \to M_2 M_3 ) &\sim
&\int\!\! d x_1 d x_2 d x_3 b_1 d b_1 b_2 d b_2 b_3 d b_3 \non &&
\hspace{-2cm}\cdot \mathrm{Tr} \left [ C(t) \Phi_B(x_1,b_1) \Phi_{M_2}(x_2,b_2)
\Phi_{M_3}(x_3, b_3) H(x_i, b_i, t) S_t(x_i)\, e^{-S(t)} \right ],
\quad \label{eq:a2}
\eeq
where $b_i$ is the conjugate space coordinate of $k_{iT}$. The large double logarithms
($\ln^2 x_i$) on the longitudinal direction are summed by the
threshold resummation, and they lead to the first Sudakov factor $S_t(x_i)$ which
smears the end-point singularities on $x_i$ \cite{li2003}.
The Sudakov resummations  of large logarithmic corrections, such as
the terms proportional to $\alpha_s\log^2[Q/{\bf k}_{\rm iT}]$ ( $Q\sim m_B$), to the $B_q$ and two final state meson
wave functions will lead to the second Sudakov factor $e^{-S(t)} = e^{-S_B(t)} \cdot
e^{-S_{M_2}(t)}\cdot e^{-S_{M_3}(t)}$. These two kinds of Sudakov factors can
together suppress the soft dynamics effectively \cite{li2003}.

In the momentum space, the light-cone wave function of $B_q$ meson can be defined as\cite{keum01-a,keum01-b,lu01:pipi},
\beq
\Phi_{B_q}(k) &=& \frac{i}{\sqrt{2 N_c}} \biggl[ (\psl + m_{B_q}) \gamma_5 \phi_{B_q}
(k) \biggr]_{\alpha\beta} \;,
\label{eq:def-wf}
\eeq
where $P$ is the momentum of the $B_q$ meson, $k$ is the momentum carried by the
light quark in $B_q$ meson, and $\phi_{B_q}$ is the corresponding distribution amplitude.

For the $B/B_s$ mesons, the distribution amplitudes  $\phi_B(x, b)$ in the $b$
space can be written as ~\cite{keum01-a,keum01-b,lu01:pipi}
\beq
\phi_{B}(x,b)&=& N_Bx^2(1-x)^2
\exp\left[-\frac{1}{2}\left(\frac{xm_B}{\omega_b}\right)^2
-\frac{\omega_b^2 b^2}{2}\right] \;,
\eeq
and
\beq
\phi_{B_s}(x,b)&=& N_{B_s} x^2(1-x)^2
\exp\left[-\frac{1}{2}\left(\frac{xm_{B_s}}{\omega_{B_s}}\right)^2
-\frac{\omega_{B_s}^2 b^2}{2}\right] \;,
\eeq
where the normalization factors $N_{B_{(s)}}$ are related to the decay constants $f_{B_{(s)}}$ through
\beq
\int_0^1 dx \phi_{B_{(s)}}(x, b=0) &=& \frac{f_{B_{(s)}}}{2 \sqrt{6}}\;.
\eeq
Here the shape parameter $\omega_b$ has been fixed at $0.40$~GeV by using the rich experimental
data on the $B$ mesons with $f_{B}= 0.19$~GeV. Correspondingly, the normalization constant $N_B$
is $91.745$. For $B_s$ meson, considering a small SU(3) symmetry breaking,
since $s$ quark is heavier than the $u$ or $d$ quark,
the momentum fraction of $s$ quark should be a little larger than that
of $u$ or $d$ quark in the $B$ mesons, we therefore adopt the shape parameter
$\omega_{B_s} = 0.50$~GeV~\cite{ali2007} with $f_{B_s} = 0.23$~GeV, then the corresponding normalization
constant is $N_{B_s} = 63.67$. In order to analyze the uncertainties of
theoretical predictions induced
by the inputs, we can vary the shape parameters $\omega_{b}$ and $\omega_{B_s}$ by
10\%, i.e., $\omega_b = 0.40 \pm 0.04$~GeV and $\omega_{B_s} = 0.50 \pm 0.05$~GeV, respectively.

For the $\pi^\pm$ and $K^\pm $ mesons, we  adopt the same set of distribution amplitudes
$\phi_{\pi,K}^A(x_i)$ and $\phi_{\pi,K}^{P,T}(x_i)$ as defined
in Refs.~\cite{ball1999,ball2006}):
\beq
\phi_{\pi(K)}^A(x) &=& \frac{3 f_{\pi(K)}}{\sqrt{6}}\, x(1-x)
\left[1 + a_1^{\pi(K)} C_1^{3/2}(t) + a_2^{\pi(K)}C_2^{3/2}(t)
+a_4^{\pi(K)}C_4^{3/2}(t)\right] \;, \label{eq:phipik-a}
\eeq
\beq
\phi^P_{\pi(K)}(x) &=& \frac{f_{\pi(K)}}{2\sqrt{2N_c}}\,
\left [ 1 +\left(30\eta_3 -\frac{5}{2}\rho_{\pi(K)}^2\right) C_2^{1/2}(t)\right.  \non
& & \left.  -\, 3\left\{ \eta_3\omega_3 +
\frac{9}{20}\rho_{\pi(K)}^2(1+6a_2^{\pi(K)}) \right\}
C_4^{1/2}(t) \right ]\;, \label{eq:phipik-p}
\\
\phi^T_{\pi(K)}(x) &=& -\frac{f_{\pi(K)}}{2\sqrt{6}}\, t\left [ 1
+ 3\left(5\eta_3 -\frac{1}{2}\eta_3\omega_3 -
\frac{7}{20}
    \rho_{\pi(K)}^2 - \frac{3}{5}\rho_{\pi(K)}^2 a_2^{\pi(K)} \right)
(5t^2-3) \right ]\;,
\label{eq:phipik-t}
\eeq
where $t=2x-1$, $\rho_{\pi(K)}=m_{\pi(K)}/m_0^{\pi(K)}$ are the mass ratios ( here
$m_0^\pi=m_\pi^2/(m_u+m_d)$ and $m_0^K=m_K^2/(m_s+m_d)$ are the chiral mass of pion and kaon),
$a_i^{\pi,K}$  are the Gegenbauer moments, while $C_n^{\nu}(t)$ are the
Gegenbauer polynomials
\beq
C_1^{3/2}(t)\, &=&  3\, t \;, \non
C_2^{1/2}(t)\, &=&\, \frac{1}{2} \left(3\, t^2-1\right), \quad
C_2^{3/2}(t)\, =\, \frac{3}{2} \left(5\, t^2-1\right), \non
C_4^{1/2}(t)\, &=& \, \frac{1}{8} \left(3-30\, t^2+35\, t^4\right), \quad
C_4^{3/2}(t) \,=\, \frac{15}{8} \left(1-14\, t^2+21\, t^4\right) \;.
\label{eq:cii}
\eeq
Under the replacement of $x\to 1-x$, only $C_1^{3/2}(t)$ will change its sign, others remain unchanged.

\section{Perturbative calculation in the pQCD approach } \label{sec:3}

In the pQCD factorization approach, the four annihilation Feynman diagrams for
$B_s \to \pi^+ \pi^-$ and $B^0 \to K^+ K^-$ decays are shown in Fig.1, where (a)
and (b) are factorizable diagrams, while (c) and (d) are the
non-factorizable ones.
The initial $\bar{b}$ and $s$($d$) quarks annihilate into $u$ and $\bar{u}$ pair,
and  then form a pair
of light mesons by hadronizing with another pair of $d\bar{d}$ ($s\bar{s}$) produced
perturbatively through the one-gluon exchange mechanism.
Besides the short-distance contributions based on one-gluon-exchange,
the $q\bar{q}$ pair can also be produced
through strong interaction in non-perturbative regime (final state interaction(FSI),
for example). FSI effects in considered decays have been assumed rather small,
we do not consider them here.

\begin{figure}[]
\vspace{-6cm}
\centerline{\epsfxsize=20 cm \epsffile{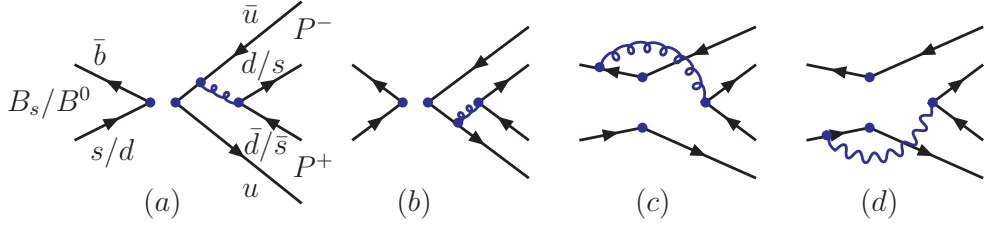}}
\vspace{-18cm}
\caption{The typical annihilation Feynman diagrams for
$B_s^0 \to \pi^+\pi^-$ and $B^0\to K^+ K^-$ decays.(a) and (b) are
factorizable diagrams; while (c) and (d) are the non-factorizable ones. }
\label{fig:fig1}
\end{figure}

We will adopt $(F^{LL},F^{LR},F^{SP})$ and $(M^{LL},M^{LR},M^{SP})$ to stand for the contributions
of the factorizable (Fig.1(a) and 1(b)) and non-factorizable (Fig.1(c) and 1(d)) annihilation diagrams
from the $(V-A)(V-A)$, $(V-A)(V+A)$ and $(S-P)(S+P)$ operators, respectively.
By making the analytic calculations we obtain the following decay amplitudes for both
$B_s^0 \to \pi^+ \pi^-$ and $B^0 \to K^+ K^-$ decays:

From the factorizable annihilation diagrams Fig.1(a) and Fig.1(b) we have
\begin{enumerate}

\item[]{(i) $(V-A)(V-A)$ operators:}
 \beq
F^{LL}&=& 16 \pi C_F M_{B_q}^2 \int_0^1 d x_{2} dx_3\, \int_{0}^{\infty} b_2 db_2 b_3
db_3\,
%\right.
\non & &
\times \left \{ \left [ x_2 \phi_2^A(x_2) \phi_3^A(x_3)
+ 2r_2 r_3 \left ( \phi_2^P(x_2) -\phi_2^T(x_2)\right ) \phi_3^P(x_3)\right.\right.\non
&& \left.\left.
+ 2r_2 r_3 x_2 \left ( \phi_2^P(x_2) +\phi_2^T(x_2)\right ) \phi_3^P(x_3) \right ] \cdot
h_{a}(x_2,x_3,b_2,b_3) \;E_a(t_a)\right.\non
&& \left.
+ \left [ (x_3-1) \phi_2^A(x_2) \phi_3^A(x_3)
-4 r_2 r_3 \phi_2^P(x_2)\phi_3^P(x_3) \right.\right.\non
&& \left.\left.
+ 2r_2 r_3 x_3 \phi_2^P(x_2) \left ( \phi_3^P(x_3)- \phi_3^T(x_3)\right) \right ] \cdot
h_b(x_2,x_3,b_2,b_3) \;E_a(t_b)\right \}, \label{eq:fll}
\eeq

\item[]{(ii) $(V-A)(V+A)$ operators:}
\beq
F^{LR} &=& F^{LL}\;, \label{eq:flr}
\eeq

\item[]{(iii) $(S-P)(S+P)$ operators:}
\beq
F^{SP}&=& 32 \pi C_F M_{B_q}^2 \int_0^1 d x_{2} dx_3\, \int_{0}^{\infty} b_2 db_2 b_3
db_3\,
%\right.
\non & &
\times \left \{ \left [ 2 r_3 \phi_2^A(x_2) \phi_3^P(x_3)
+ r_2 x_2 \left ( \phi_2^P(x_2) -\phi_2^T(x_2)\right ) \phi_3^A(x_3)\right ]\right.\non
&& \left.
\cdot h_a(x_2,x_3,b_2,b_3) \cdot E_a(t_a)\right.\non
&& \left.
\left [ 2 r_2 \phi_2^P(x_2) \phi_3^A(x_3)
+(1-x_3)r_3 \phi_2^A(x_2) \left ( \phi_3^P(x_3)+ \phi_3^T(x_3)\right) \right ] \right. \non
&& \left.
\cdot h_b(x_2,x_3,b_2,b_3) \cdot E_a(t_b)\right \}, \label{eq:fsp}
\eeq
\end{enumerate}
where $r_2=m_2/m_{B_q}$, $r_3=m_3/m_{B_q}$ with $q=(d,s)$ for $B_d^0$ or $B_s^0$ decays,
and $C_F=4/3$ is a color factor. The explicit expressions for the convolution functions
$E_a(t_{a,b})$, the hard scales $t_{a,b}$, and the hard functions $h_{a,b}(x_i,b_i)$ can
be found for example in Ref.~\cite{wang2006a,xiao2008b}.

From the non-factorizable annihilation diagrams Fig.1(c) and Fig.1(d) we have
\beq
M^{LL}&=& \frac{64}{\sqrt{6}} \pi C_F M_{B_q}^2 \int_0^1 dx_1 d x_{2} dx_3\,
\int_{0}^{\infty} b_1 db_1 b_2 db_2 \,\phi_{B_q}(x_1,b_1)
%\right.
\non & &
\times \left \{ \left [ (1-x_3) \phi_2^A(x_2) \phi_3^A(x_3)
+ (1-x_3)r_2 r_3 \left ( \phi_2^P(x_2) +\phi_2^T(x_2)\right ) \left (\phi_3^P(x_3)-\phi_3^T(x_3)\right)\right.\right.\non
&& \left.\left.
+ x_2r_2 r_3 \left ( \phi_2^P(x_2) -\phi_2^T(x_2)\right ) \left ( \phi_3^P(x_3) +\phi_3^T(x_3)\right )
\right ] \cdot
h_c(x_1,x_2,x_3,b_1,b_2) \;E_c(t_c)\right.\non
&& \left.
+ \left [ -x_2 \phi_2^A(x_2) \phi_3^A(x_3)
-4 r_2 r_3 \phi_2^P(x_2)\phi_3^P(x_3) \right.\right.\non
&& \left.\left.
+  (1-x_2)r_2 r_3 \left ( \phi_2^P(x_2)+ \phi_2^T(x_2)\right)\left ( \phi_3^P(x_3)- \phi_3^T(x_3)\right)
\right.\right.\non
&& \left.\left.
+ x_3 r_2 r_3  \left ( \phi_2^P(x_2)- \phi_2^T(x_2)\right)\left ( \phi_3^P(x_3)+ \phi_3^T(x_3)\right)
 \right ]
 \right.\non
&& \left.
\hspace{1cm} \cdot
h_d(x_1,x_2,x_3,b_1,b_2) \;E_c(t_d)\right \}, \label{eq:mll}
\eeq
\beq
M^{LR}&=& \frac{64}{\sqrt{6}} \pi C_F M_{B_q}^2 \int_0^1 dx_1 d x_{2} dx_3\,
\int_{0}^{\infty} b_1 db_1 b_2 db_2 \,\phi_{B_q}(x_1,b_1)
%\right.
\non & &
\times \left \{ \left [ x_2 r_2  \left ( \phi_2^P(x_2) +\phi_2^T(x_2)\right ) \phi_3^A(x_3)
- (1-x_3) r_3 \phi_2^A(x_2)\left ( \phi_3^P(x_3) -\phi_3^T(x_3)\right ) \right]\right.\non
&& \left.
\hspace{1cm}\cdot h_c(x_1,x_2,x_3,b_1,b_2) \;E_c(t_c)\right.\non
&& \left.
+ \left [ (2-x_2)r_2  \left ( \phi_2^P(x_2) +\phi_2^T(x_2)\right ) \phi_3^A(x_3)
- (1+x_3) r_3 \phi_2^A(x_2)\left ( \phi_3^P(x_3) -\phi_3^T(x_3)\right ) \right]\right.\non
&& \left.
\hspace{1cm}\cdot h_d(x_1,x_2,x_3,b_1,b_2) \;E_c(t_d)\right \}, \label{eq:mlr}
\eeq
%%%%
\beq
M^{SP}&=& \frac{64}{\sqrt{6}} \pi C_F M_{B_q}^2 \int_0^1 dx_1 d x_{2} dx_3\,
\int_{0}^{\infty} b_1 db_1 b_2 db_2 \,\phi_{B_q}(x_1,b_1)
%\right.
\non & &
\times \left \{ \left [ x_2  \phi_2^A(x_2) \phi_3^A(x_3)
+ x_2 r_2 r_3 \left ( \phi_2^P(x_2) +\phi_2^T(x_2)\right ) \left ( \phi_3^P(x_3) -\phi_3^T(x_3)\right )
\right.\right.\non
&& \left.\left.
+ (1-x_3) r_2 r_3  \left ( \phi_2^P(x_2) -\phi_2^T(x_2)\right ) \left ( \phi_3^P(x_3) +\phi_3^T(x_3)\right )
\right]\right.\non
&& \left.
\hspace{1cm}\cdot h_c(x_1,x_2,x_3,b_1,b_2) \;E_c(t_c)\right.\non
&& \left.
+ \left [ -(1-x_3) \phi_2^A(x_2) \phi_3^A(x_3) -4 r_2 r_3  \phi_2^P(x_2) \phi_3^P(x_3)
\right.\right.\non
&& \left.\left.
+ x_3 r_2 r_3  \left ( \phi_2^P(x_2) +\phi_2^T(x_2)\right ) \left ( \phi_3^P(x_3) -\phi_3^T(x_3)\right )
\right.\right.\non
&& \left.\left.
+ (1-x_2) r_2 r_3  \left ( \phi_2^P(x_2) -\phi_2^T(x_2)\right ) \left ( \phi_3^P(x_3) +\phi_3^T(x_3)\right )
\right]\right.\non
&& \left.
\hspace{1cm}\cdot h_d(x_1,x_2,x_3,b_1,b_2) \;E_c(t_d)\right \}, \label{eq:msp}
\eeq
where $r_{2,3}$ and $C_F$ are defined in the same way as in Eqs.(\ref{eq:fll}-\ref{eq:fsp}).
Again, the explicit expressions of the functions $E_c(t_{c,d})$ and $h_{c,d}$, and the hard scales $t_{c,d}$
can be found in Ref.~\cite{wang2006a,xiao2008b}.

Because of the isospin symmetry, the contributions to both $B_s^0 \to \pi^+ \pi^-$ and $B^0 \to K^+ K^-$ decays
from the factorizable annihilation diagrams Fig.1(a) and Fig.1(b) cancel each other.
The total decay amplitudes for the considered decays are therefore written as:
\beq
{\cal A}(B_s^0 \to \pi^+ \pi^-) &=& V_{ub}^* V_{us}\; C_2 M^{LL}\non
&&  \hspace{-3cm}
- V_{tb}^* V_{ts}\; \left \{
\left [ C_4+C_6-\frac{1}{2}C_8 +C_{10} \right ]\; M^{LL}
+ \left [ C_4+C_6+ C_8 -\frac{1}{2}C_{10}\right ]\; M^{SP}  \right\},\; \label{eq:abspipi}
\eeq
\beq
{\cal A}(B_d^0 \to K^+ K^-) &=&  V_{ub}^* V_{ud}\; C_2 M^{LL}\non
&&  \hspace{-3cm}
- V_{tb}^* V_{td}\; \left \{ \left [ C_4+C_6-\frac{1}{2}C_8 +C_{10} \right ]\; M^{LL}
+ \left [ C_4+C_6+ C_8 -\frac{1}{2}C_{10}\right ]\; M^{SP}  \right\}.\; \label{eq:abkk}
\eeq
The expression of decay amplitude in Eq.~(\ref{eq:abspipi}) is  equivalent with those as given  in
Refs.\cite{liy2004,ali2007} by a proper transformation between $M^{LL}$ and $M^{SP}$.

\section{Numerical Results and Discussions}\label{sec:4}

Now it is straightforward to calculate the CP-averaged branching ratios and
CP-violating asymmetries
for the two considered decays. In numerical calculations, central values of
the input parameters will be
used implicitly unless otherwise stated. The QCD scale~({\rm GeV}),
masses~({\rm GeV}), decay constants~({\rm GeV}),
and $B_q$ meson lifetime ({\rm ps}) being used are the following\cite{pdg2010}
\beq
\Lambda_{QCD}&=& 0.25\; , \quad m_W = 80.41\;, \quad m_{B^0} = 5.2795\;,
\quad  M_{B_s}= 5.3663\;; \non
m_\pi&=&0.14, \quad m_K=0.494, \quad f_\pi= 0.13, \quad f_K=0.16, \non
\tau_{B^0} &=& 1.525\;ps,\quad \tau_{B_s} = 1.472\;ps.
\label{eq:mass}
\eeq
As for the CKM matrix elements, we use\cite{pdg2010}
\beq
\lambda &=& 0.2253\pm 0.0007, \quad A= 0.808^{+0.022}_{-0.015} \quad \bar{\rho} = 0.132^{+0.022}_{-0.014},
\quad \bar{\eta}= 0.341\pm 0.013.
\eeq
For the Gegenbauer moments and other relevant input parameters, based on the works of \cite{ball1999,ball2006}, we use
\beq
a_1^\pi &=& 0, \quad a_1^K = 0.06\pm 0.03, \quad a_2^{\pi}=0.35\pm 0.15, \quad a_2^K=0.25\pm 0.10,
\non
a_4^{\pi} &=& -0.015, \quad a_4^K=0, \quad \rho_\pi = m_\pi/m_0^\pi, \quad \rho_K = m_K/m_0^K,\non
\eta_3&=&0.015\pm 0.005, \quad \omega_3=-3.0\pm 1.0 \label{eq:pa-1}
\eeq
with the chiral mass $m_0^\pi=1.4 \pm 0.1$ GeV, and $m_0^K=1.9 \pm 0.2$ GeV. In order to check the theoretical errors
induced by the uncertainty of the Gegenbauer moments we vary $a_1^K,$ $a_2^{\pi, K}$, $\eta_3$ and $\omega_3$ in the range of
$a_1^K=0.06 \pm 0.03$, $a_2^\pi=0.35\pm 0.15$, $a_2^K=0.25\pm 0.10$, $\eta_3=0.015 \pm 0.005$ and $\omega_3=-3.0 \pm 1.0$.

From the decay amplitudes, it is easy to write down the corresponding branching ratio:
\beq
Br(B\to P P) = \frac{G_F^2 m_B^3}{128\pi^2}\; \tau_B\left |{\cal A}(B \to PP)\right |^2,
\eeq
where ${\cal A}(B \to PP)$ is the decay amplitude as defined in Eqs.~(27,28).

By using the analytic expressions for the complete decay amplitudes and the
input parameters, we calculate the branching ratios  and CP-violating asymmetries for both considered decay modes.
The numerical results are the following:
\beq
Br(B_s^0 \to \pi^+ \pi^-) &=& \left ( 5.10^{+1.96}_{-1.68}(a_2^\pi) ^{+0.25+1.05+0.29}_{-0.19 -0.83-0.20}\right)\times 10^{-7}, \label{eq:brpipi}\\
{\cal A}_{CP}(B_s^0 \to \pi^+ \pi^-) &=& \left ( -2.3^{+0.0}_{-0.3}(a_2^\pi) ^{+ 0.3+0.1+0.1}_{-0.2-0.2-0.1} \right )\%, \label{eq:acppipi}\\
Br(B_d^0 \to K^+ K^-) &=& \left ( 1.56^{+0.44}_{-0.42}(a_2^K) ^{+0.23 +0.22+0.13}_{-0.22-0.19-0.09}\right)\times 10^{-7}, \label{eq:brkk}\\
{\cal A}_{CP}(B_d^0 \to K^+ K^-) &=& \left ( 18.9^{+0.2}_{-1.9}(a_2^K)^{+1.4+0.1+0.8}_{-2.2-1.4-1.1}\right )\%,
\label{eq:acpkk}
\eeq
where the first error comes from the theoretical uncertainty of the Gegenbauer moments
$a_2^\pi=0.35\pm 0.15$ and $a_2^K=0.25\pm 0.10$, the small  theoretical errors due to the variations of 
$a_1^K=0.06\pm 0.03$, $\eta_3=0.015\pm 0.005$ and $\omega_3=-3.0\pm 1.0$ is shown as the second error, 
the third error includes the uncertainties
induced by the parameter $\omega_b=0.40 \pm 0.04$ GeV and $\omega_{B_s}=0.50\pm 0.05$ GeV, as well as the uncertainties of
$m_0^\pi=1.4\pm 0.1$ GeV and $m_0^K=1.9\pm 0.2$ GeV, and the last error comes from the uncertainties
of the relevant CKM elements. It is easy to see that the uncertainties from $a_2^{\pi,K}$, $\omega_b$ and $\omega_{B_s}$
dominate the theoretical error.

For $B_s^0 \to \pi^+ \pi^-$ decay, the pQCD prediction for its branching ratio
in Eq.~(\ref{eq:brpipi}) agree very
well with the measured results from CDF and LHCb collaboration
\cite{cdf-2011a,lhcb-2011a} as shown in  Eqs.~(\ref{eq:exppi}).
This pQCD prediction also agree very well with the previous pQCD predictions
as given in Refs.~\cite{liy2004,ali2007,xiao2008}.
The analytical results for the decay amplitudes obtained in this paper are
consistent with those as given in Refs.\cite{liy2004,ali2007,xiao2008}.
The small difference in numerical pQCD predictions comes from the difference
of the input parameters being used in different works.

For $B_d^0 \to K^+ K^-$ decay, fortunately, the pQCD prediction for its branching
ratio  in Eq.~(\ref{eq:brkk}) agrees well with the measured results from CDF and LHCb collaboration
\cite{cdf-2011a,lhcb-2011a} as shown in  Eqs.~(\ref{eq:expk}).

It is easy to see that the new pQCD prediction in Eq.~(\ref{eq:brkk}) is much larger
than  the previous pQCD prediction as given in  Ref.~\cite{chen2001}.
In order to find the reason for the large difference, we checked the relevant analytical expressions
as given in Ref.~\cite{chen2001} and found that those analytical results are
consistent with our results after proper transformation: $x\to 1-x$.
The large numerical difference between two pQCD predictions comes from the fact that
(a) the distribution amplitudes of the kaon meson used by Chen and Li \cite{chen2001}
are very different from those used in this paper; and (b) some improved Gegenbauer moments
as given in Ref.~\cite{ball2006} are used in this paper.

In Ref.\cite{chen2001}, only the axial-vector and pseudo-scalar kaon wave functions
$\phi_K(x)$ and $\phi_K^\prime(x)$ were considered:
\beq
\phi_K(x) &=& \frac{3f_K}{\sqrt{6}}\, x(1-x)\left \{ 1 +0.51(1-2x)+ 0.3 \left [5(1-2x)^2-1\right ] \right \}, \label{eq:chen1} \\
\phi^\prime_K(x) &=& \frac{3 f_K}{\sqrt{6}}\, x(1-x), \label{eq:chen2}
\eeq
In this paper, however, besides the leading twist-2 $\phi_K^A(x)$ (i.e. the axial-vector
$\phi_K(x)$ in Ref.~\cite{chen2001} ), we also take into account the twist-3 contributions
from both
$\phi_K^P$ and $\phi_K^T$ simultaneously. Based on the analytical expressions
as given in Eqs.~(\ref{eq:phipik-a}-\ref{eq:phipik-t}), one can obtain the numerical expressions for
$\phi_K^A(x), \phi_K^P(x)$ and $\phi_K^T(x)$:
\beq
\phi_K^A(x) &=& \frac{3f_K}{\sqrt{6}}\, x(1-x)\left \{ 1 -0.18(1-2x) + 0.375 \left [5(1-2x)^2-1\right] \right\}, \label{eq:phi-0a} \\
\phi^P_K(x) &=& \frac{f_K}{2\sqrt{6}}\, \left \{ 1 + 0.282 (1-6x+6x^2 )
-0.012 \left [3-30(2x-1)^2+35(2x-1)^4 \right ] \right \},\ \  \label{eq:phi-0b} \\
\phi^T_K(x) &=&  -\frac{f_K}{2\sqrt{6}}\, (2x-1)\left [ 1 + 0.55 \left (1-10x + 10x^2\right ) \right ],\label{eq:phi-0c}
\eeq
by using the central values of the relevant input parameters $a_{1,2,4}^K, \rho_k, \eta_3$ and $\omega_3$, etc,  as
given in Eqs.(\ref{eq:mass},\ref{eq:pa-1}).

For the leading twist-2 axial-vector wave function, the $\phi_K^A(x)$ we used is in the same form as $\phi_K(x)$ being used
in Ref.~\cite{chen2001}. The difference of the coefficients of the second and third term comes from the variation
of the values of the corresponding Gegenbauer moments $(a_1^K, a_2^K)$: $(a_1^K, a_2^K)=(0.17,0.20)$ in Ref.~\cite{chen2001}, while
$(a_1^K, a_2^K)=(0.06,0.25)$ in this paper, based on recent improvements made in Ref.~\cite{ball2006}.
The difference of the sign of the second term in $\phi_K^A(x)$ is resulted from the different assignment for the
momentum fraction $x$ in Ref.~\cite{chen2001} and in this paper: We here use $x$ to denote the momentum fraction
of $s/\bar{s}$ quark in the $K^\pm$ meson, instead of the $u/\bar{u}$ quark as assigned in Ref.~\cite{chen2001}.
The Gegenbauer polynomial $C_1^{3/2}(t)=3 t$ in Eq.~(38) will change its sign under the transformation $x \to 1-x$.

In Ref.~\cite{chen2001}, the authors took $\phi^\prime_K(x) = \frac{3}{\sqrt{6}}\,f_K x(1-x)$ as the pseudo-scalar
kaon wave function, which was "determined from the data of the $B\to K \pi$ decays" by Chen and Li,
instead of the ordinary $\phi_K^P(x)$ as derived from the QCD sum rule \cite{ball1999,ball2006} and shown in Eqs.~(39).
(for more details of the derivation of $\phi_K^\prime(x)$, see Sec.IV of Ref.~(keum01-b)).
The $\phi^\prime_K(x)$ in Ref.~\cite{chen2001} is just the first and leading term of the twist-2 part $\phi_K^A(x)$
and is very different from commonly used $\phi_K^P$.

In Ref.~\cite{chen2001}, the term $\phi_K^T(x)$ was absent. All differences in the relevant wave functions
being used in Ref.~\cite{chen2001} and in this paper lead to the large difference between the pQCD predictions
for the branching ratio $Br(B^0 \to K^+ K^-)$ as presented in Ref.~\cite{chen2001} and in this paper.

Explicit numerical examinations also show that the leading twist-2 $\phi_K^A$ provide the dominant contribution to
the magnitude of the decay amplitudes and consequently branching ratio $Br(B^0 \to K^+ K^-)$:
\begin{enumerate}
\item
When all three terms $\phi_K^{A,P,T}$, or only  the leading twist-2 term $\phi_K^A(x)$, are taken into account, we find
numerically
\beq
{\cal A}(B^0 \to K^+ K^-)&=& \left\{ \begin{array}{l l}
(-0.31 - 2.2\;{\rm I})\times 10^{-5}, & ({\rm \phi_K^A(x) \ \ only});  \\
(-0.82 - 3.6\;{\rm I})\times 10^{-5}, & ( {\rm All \ \ three\ \ terms});  \\ \end{array} \right.  \\
Br(B^0 \to K^+ K^-)&=& \left\{ \begin{array}{l l}
0.55 \times 10^{-7}, & ({\rm \phi_K^A(x) \ \ only});  \\
1.56 \times 10^{-7}, & ({\rm All \ \ three\ \ terms});  \\ \end{array} \right. .
\eeq

\item
If only  the twist-3 term $\phi_K^P(x)$, $\phi_K^T$ or both of them are taken into account, we find numerically
\beq
{\cal A}(B^0 \to K^+ K^-)&=& \left\{ \begin{array}{l l}
(-0.61 - 0.55\; {\rm I})\times 10^{-5}, &  ( {\rm \phi_K^P(x) \ \ only});  \\
(0.06  - 0.27\; {\rm I})\times 10^{-5} , & ( {\rm \phi_K^T(x) \ \ only});  \\
\end{array} \right.  \\
Br(B^0 \to K^+ K^-)&=&  \left\{ \begin{array}{l l}
0.08 \times 10^{-7}, & ( {\rm\phi_K^P(x) \ \ only});  \\
0.01 \times 10^{-7}, & ( {\rm\phi_K^T(x) \ \ only});  \\ \end{array} \right.
\eeq
\end{enumerate}
It is straightforward to see from the above numerical results that
\begin{enumerate}
\item
The leading twist-2 term $\phi_K^A(x)$ provide the dominant contribution to the decay amplitude:
${\cal A}=(-0.31-2.2\; I ) \times 10^{-5}$ if only $\phi_K^A(x)$ is taken into account,
while ${\cal A}=(-0.61-0.55\; I ) \times 10^{-5}$ ( ${\cal A}=(-0.06-0.27\; I ) \times 10^{-5}$ ) if only
$\phi_K^P(x)$ ( $\phi_K^T$) is taken into account.
For the branching ratio, its size would be $10^{-7}$, $10^{-8}$ or $10^{-9}$ if only the term $\phi_K^A(x)$,
$\phi_K^P(x)$ or  $\phi_K^T(x)$ contribute.

\item
The enhancements due to the constructive interference between the three parts also play an important role in
producing a large branching ratio $Br(B^0 \to K^+ K^-)$.
One  can see that the contributions to the decay amplitude ${\cal A}$ from the three terms interfere constructively,
which finally leads to a large branching ratio $Br(B^0 \to K^+ K^-) = 1.56 \times 10^{-6}$, partially due to the further
magnifying effects since the branching ratio is proportional to the  module square of the decay amplitude ${\cal A}$.

\end{enumerate}

As for the CP-violating asymmetry for the considered decays, ${\cal A}_{CP}
(B_s^0 \to \pi^+\pi^-)$
is very small,only about two percent and therefore hardly to be detected even at the LHCb.
For $B_d^0 \to K^+ K^-$ decay, however,
its ${\cal A}_{CP}$ is relatively large, around $19\%$, and may be detected
at the LHCb experiment or future Super-B factory experiments.

In summary, by employing the pQCD factorization approach, we here recalculated
the branching ratios and CP-violating asymmetries of the pure annihilation decays
$B_s^0 \to \pi^+\pi^-$ and $B^0\to K^+ K^-$ with the usage of the wave functions based on the QCD sum
rule \cite{ball1999,ball2006}
and the improved Gegenbauer moments \cite{ball2006}.
By numerical calculations and phenomenological analysis we found the following results:
(a) one can provide a consistent pQCD interpretation for both the
measured $Br(B_s^0 \to \pi^+ \pi^-)$ and $Br(B_d^0 \to K^+ K^-)$ simultaneously;
(b) the pQCD predictions for $Br(B_s^0 \to \pi^+\pi^- )$ obtained
by different authors are well consistent with each other within one
standard deviation;
(c) our new pQCD prediction for $Br(B_d^0 \to K^+K^- )$ agrees well with
the measured values from CDF and LHCb Collaboration;
and (d) the CP-violating asymmetry ${\cal A}_{CP}(B_s^0 \to \pi^+\pi^-)
\approx -2.3\%$, may be too small to be detected even at LHCb experiment;
(e) ${\cal A}_{CP}(B_d^0 \to K^+K^-) \approx 19\%$, which is large and may be
detected at the LHCb and future super-B factory experiments.

\begin{acknowledgments}

This work is supported by the National Natural Science Foundation of China under Grant No. 10975074 and 10735080.

\end{acknowledgments}

%%%%%%%%%%%%%%%%%%%%%%%%%%%%%%%%%%%%%%%%%%%%%%%%%%%%%%%%%%%%%%%%%%%%%%%%%%%%%%%%%%%%%%%%%%%%%%5
%                                 reference
%%%%%%%%%%%%%%%%%%%%%%%%%%%%%%%%%%%%%%%%%%%%%%%%%%%%%%%%%%%%%%%%%%%%%%%%%%%%%%%%%%%%%%%%%%%%%%%%%


\begin{thebibliography}{99}

\bibitem{chen2001}
C.H.~Chen and H.N.~Li,  \prd {\bf 63}, 014003 (2000).
% FSI and $B \to KK$ decays in perturbative QCD

\bibitem{liy2004}
Y.~Li, C.D.~L\"u, Z.J.~Xiao, and  X.Q.~Yu,  \prd {\bf 70}, 034009 (2004).
%% Branching ratio and CP asymmetry of B(s) ---> pi+ pi- decays in the pQCD approach.

\bibitem{ali2007}
A.~Ali, G.~Kramer, Y.~Li, C.D.~L\"u, Y.L.~Shen, W.~Wang, and Y.M.~Wang,
\prd {\bf 76}, 074018 (2007);
%" Charmless nonleptonic Bs decays to PP, PV, and VV final states in the pQCD approach"

\bibitem{xiao2008}
J.~Liu, R.~Zhou and Z.J.~Xiao, arXiv:0812.2312v1 [hep-ph].
%" $B_s\to PP$ decays and the NLO contributions in the pQCD approach"

\bibitem{bn675}
M.~Beneke and M.~Neubert, \npb {\bf 675}, 333 (2003).

\bibitem{duprd68}
J.F.~Sun, G.H.~Zhu and D.S.~Du, \prd {\bf 68}, 054003 (2003).

\bibitem{yang2005}
Y.D.~Yang, F.~Su, G.R. Lu and H.J. Hao, \epjc {\bf 44}, 243 (2005).
%% Revisiting the annihilation decay $\bar{B}_s \to \pi^+ \pi^-$.

\bibitem{cheng1}
H.Y.~Cheng and C.K.~ Chua, \prd {\bf 80}, 114008 (2009).
%% Revisiting Charmless Hadronic B(u,d) Decays in QCD Factorization.

\bibitem{cheng2}
H.Y.~Cheng and C.K.~ Chua, \prd {\bf 80}, 114026 (2009).
%% QCD Factorization for Charmless Hadronic B_s Decays Revisited.
%% arXiv:0910.5237 [hep-ph]

\bibitem{zhu2011}  %%  10
G.H.~Zhu, \plb {\bf 702}, 408 (2011).
%% Implications of the recent measurement of pure annihilation $B_s \to \pi^+ \pi^-$
%% decays in QCD factorization;
%% arXiv:1106.4709V2 [hep-ph]

\bibitem{cdf-2011a}
F.~Ruffini, CDF Collaboration, talk given at the Flavor Physics and CP
violation 2011, May 23-27, Israel; arXiv:1107.5760[hep-ex];
%% Measurements of B meson decay rates and CP-violating asymmetries;
M.J.~Morello {\it et al.,}, (CDF Collaboration), CDF public note 10498 (2011);
%% First evidence of $B^0_s\to \pi^+ \pi^-$ decays.
T.~Aaltonen {\it et al.,}, (CDF Collaboration), arXiv:1111.0485v1 [hep-ex].
%% Evidence for the charmless annihilation decay mode $B^0_s\to \pi^+ \pi^-$.

\bibitem{lhcb-2011a}
A.~Powell, LHCb Collaboration, talk given at PANIC 2011, MIT,  July 2011;
%% Hadronic B decays at LHCb;
V.~Vagnoni, LHCb Collaboration, LHCb-CONF-2011-042, Sept. 20, 2011.
%% Charmless charged two-body B decays at LHCb with 2011 data.

\bibitem{bbns1999}
M.~Beneke, G.~Buchalla, M.~Neubert, and C.T. Sachrajda, \prl {\bf 83}, 1914 (1999);
\npb {\bf 591}, 313 (2000).

\bibitem{keum01-a}
Y.Y.~Keum, H.N.~Li and A.I.~Sanda,  \plb {\bf 504}, 6 (2001).
%``Fat penguins and imaginary penguins in perturbative QCD,''

\bibitem{keum01-b}  %% 15
Y.Y.~Keum, H.N.~Li and A.I.~Sanda,  \prd {\bf 63}, 054008 (2001).
%``Penguin enhancement and B --> K pi decays in perturbative QCD,''

\bibitem{lu01:pipi}
C.D.~L\"u, K.~Ukai and M.Z.~Yang,  \prd {\bf 63}, 074009 (2001).
%``Branching ratio and {\IT CP} violation of B --> pi pi decays in perturbative  QCD
%approach,''

\bibitem{li2003}
H.N.~Li, Prog. Part. $\&$ Nucl. Phys. {\bf 51},  85 (2003),  and
reference therein.
%[QCD Aspects of Exclusive B Meson Decays]


\bibitem{wang2006a}
H.S.~Wang, X.~Liu, Z.J.~Xiao, L.B.~Guo, and C.D.~L\"u, \npb {\bf 738}, 243 (2006);
X.~Liu, H.S.~Wang, Z.J.~Xiao, L.B.~Guo, and C.D.~L\"u, \prd {\bf 73}, 074002 (2006).

\bibitem{buras96}
G.~Buchalla, A.J.~Buras, and M.E.~Lautenbacher, \rmp {\bf 68}, 1215 (1996)

\bibitem{xiao2008b}
Z.J.~Xiao, Z.Q.~Zhang, X.~Liu, and L.B.~Guo, \prd {\bf 78}, 114001 (2008).

\bibitem{buras697}
A.J.~Buras, R.~Fleischer, S.~Recksiegel and F.~Schwab, \npb {\bf 697}, 133 (2004).


\bibitem{ball1999}
P.~Ball, \jhep 9809, 005 (1998); \jhep 9901, 010 (1999).

\bibitem{ball2006}
P.~Ball and R. Zwicky, \prd {\bf 71}, 014015 (2005);
P.~Ball, V.M.~Braun, and A.~Lenz, \jhep  {\bf 0605} (2006) 004.

\bibitem{pdg2010}
K.~Nakamura {\it et al.,} (Particle Data Group), \jpg {\bf 37}, 075021 (2010).

\end{thebibliography}
\end{document}